# Friction of the microscale silica under various behaviors of the shape and the orientation of the coarse-grained particle in adaptive smoothed particle hydrodynamics

Le Van Sang[1,**]  Natsuko Sugimura[1,3] · Hitoshi Washizu[1,2*]

[1]Graduate School of Simulation Studies, University of Hyogo - Kobe, Hyogo 650-0047, Japan

[2]Elements Strategy Initiative for Catalysts and Batteries (ESICB), Kyoto University - 1-30 Goryo-Ohara, Nishikyo-ku, Kyoto 615-8245, Japan

[3]Faculty of Engineering, Tokyo City University - 1-28-1 Tamazutsumi, Setagaya, Tokyo 158-8557, Japan

[*]Email: h@washizu.org

[**]Email: levansang82@gmail.com

**Abstract**   The paper investigates dry sliding friction of the coarse-grained microscale α-SiO$_2$ oxide. Adaptive smoothed particle approach is used to consider various shapes and orientations of the particles. It is found that because of the stable system the friction characteristics almost do not depend on the shape and the orientation of the particle. The friction coefficient of 0.1376 observed in the present work is in accordance with that founded in previously experimental reports. The friction coefficient steady maintains in the applied load range of 5 – 80 µN, showing a very slightly linear drop from 0.1379 to 0.1341 in this load range. This observation is also consistent with the applied load – friction coefficient relationship mentioned in previously experimental studies. For the sinusoidal rough contacts, at a given amplitude the friction coefficient almost does not depend on the wavelength while at a given wavelength it linearly increases with the amplitude.

**Keywords**    Sliding friction · Load-friction relationship · Silica contact · ASPH simulation · Sinusoidal roughness

**1 Introduction**

Most numerical calculations or simulations have considered atoms or particles as the interaction points. In such the works, influences of shape of particle on the observed quantities are ignored. Recently, some research groups developed the interaction potentials such as the Gay-Bern anisotropic Lennard-Jones potential [1, 2] and the RE-squared anisotropic potential [1, 3] to model interaction between ellipsoidal particles or an ellipsoidal particle and a spherical particle. In other branch, the discrete element modeling (DEM) method proposed particle as the solid sphere of a given radius [4]. These works have brought out significant contributions for studies of coarse-grained (CG) particle that is usually modeled in various shapes since the particle is yielded by lumping a group of atoms. However, most studies in the past considered combination of the anisotropic potential and a CG model in molecular dynamics simulations in which system is put in limit of atomistic scale or nanoscale. The smoothed particle hydrodynamics (SPH) or adaptive smoothed particle hydrodynamics (ASPH) approach can be utilized to employ with system of microscale. While the SPH presents isotropic interaction of particle, the ASPH presents anisotropic one of particle. Therefore, in order to investigate effects of particle shape on the pointed properties of a microscale system one should carry out ASPH estimations of CG system. The ASPH also showed better the time evolution of the azimuthally averaged radial density profiles [5] and the high strain hydrodynamic problems [6] than the SPH did.

The present work investigates dry sliding friction of the CG α-$SiO_2$ oxide of microscale by the ASPH simulations. Up to now, we have not ever seen studies of sliding friction employing in this approach. Additionally, friction, adhesion and wear properties of $SiO_2$ oxide should extensively be investigated at nanoscale because of its common applications in multilayer semiconductor devices and at macroscale due to its dominant occupancy in rocks whose instable movement is related to the earthquake phenomenon. Different contributions to friction of $SiO_2$ oxide have been reported. Volokitin showed both the thermal and quantum contributions to the Casimir frictional drag force between a $SiO_2$ tip and a $SiO_2$ substrate or a graphene-covered $SiO_2$ substrate [7]. Li et al. found dependence of static friction between amorphous silica surfaces on a varying number of interfacial siloxane (Si–O–Si) bridges by molecular dynamics simulations [8]. There are also discussions of friction of quartz related to earthquake phenomenon [9, 10]. As an implementation for understandings of friction of $SiO_2$ oxide, this paper focuses on dry sliding friction study of the CG α-$SiO_2$/α-$SiO_2$ contact of microscale and monitors influences of shape and orientation of CG particle modeled by anisotropy of the kernel in the ASPH simulations on friction of the system. Effects of the sinusoidal rough contacts on friction are also considered by analyzing dependence of friction on the amplitude and the wavelength of the sinusoidal roughness.

## 2 Model and calculations

### 2.1 Adaptive smoothed particle hydrodynamics

A main difference between the SPH and the ASPH comes from consideration of the kernel function. A smoothed scalar length ($h$) of the kernel in the SPH is replaced by a second order symmetric tensor ($G$) in the ASPH. The time evolution of the density, velocity and internal energy of the particle in both is presented by the following equations

$$\frac{d\rho_i}{dt} = \sum_{j=1}^{N} m_j \left( \vec{v}_j - \vec{v}_i \right) \vec{\nabla}_i W\left( \vec{\eta}_i \right), \tag{1}$$

$$\frac{dv_i^\alpha}{dt} = \sum_{j=1}^{N} m_j \left( \frac{\sigma_i^{\alpha\beta}}{\rho_i^2} + \frac{\sigma_j^{\alpha\beta}}{\rho_j^2} + \Pi_{ij} \right) \nabla_i^\beta W\left( \vec{\eta}_i \right), \tag{2}$$

$$\frac{du_i}{dt} = \frac{1}{2} \sum_{j=1}^{N} m_j \left( \frac{\sigma_i^{\alpha\beta}}{\rho_i^2} + \frac{\sigma_j^{\alpha\beta}}{\rho_j^2} + \Pi_{ij} \right) \left( v_j^\alpha - v_i^\alpha \right) \nabla_i^\beta W\left( \vec{\eta}_i \right), \tag{3}$$

where $\rho$, $v$, $u$ and $m$ are density, velocity, internal energy and mass of the particle, respectively; t is time; $\alpha, \beta \equiv x, y, z$; $W$, $\Pi$ and $\sigma$ are kernel function, artificial viscosity function and stress tensor, respectively; and $\vec{\eta}_i = \vec{r}_{ij} / h_i$ in the SPH and $\vec{\eta}_i = G_i \vec{r}_{ij}$ in the ASPH, $\vec{r}_{ij} = \vec{r}_i - \vec{r}_j$ is relative position vector of the two particles $i$ and $j$. In all the calculations, $W(\vec{\eta}_i)$ and $\vec{\nabla}W(\vec{\eta}_i)$ are replaced by $W_{ij} = 0.5\left[W(\vec{\eta}_i) + W(\vec{\eta}_j)\right]$ and $\vec{\nabla}W_{ij} = 0.5\left[\vec{\nabla}(\vec{\eta}_i) + \vec{\nabla}W(\vec{\eta}_j)\right]$, respectively. The $G$ tensor evolves in time as follows

$$\frac{dG}{dt} = \Gamma^G G - G\sigma^G, \tag{4}$$

where the second order tensors $\sigma^G$ and $\Gamma^G$ are determined as follows

$$\sigma^G = \begin{pmatrix} \sigma_{11}^G & \sigma_{12}^G & \sigma_{13}^G \\ \sigma_{21}^G & \sigma_{22}^G & \sigma_{23}^G \\ \sigma_{31}^G & \sigma_{32}^G & \sigma_{33}^G \end{pmatrix} = \begin{pmatrix} \partial v_x/\partial x & \partial v_x/\partial y & \partial v_x/\partial z \\ \partial v_y/\partial x & \partial v_y/\partial y & \partial v_y/\partial z \\ \partial v_z/\partial x & \partial v_z/\partial y & \partial v_z/\partial z \end{pmatrix}, \tag{5}$$

$$\Gamma_G = \begin{pmatrix} \Gamma_{11}^G & \Gamma_{12}^G & \Gamma_{13}^G \\ \Gamma_{12}^G & \Gamma_{22}^G & \Gamma_{23}^G \\ \Gamma_{31}^G & \Gamma_{32}^G & \Gamma_{33}^G \end{pmatrix} = \begin{pmatrix} 0 & \Gamma_{12}^G & -\Gamma_{31}^G \\ -\Gamma_{12}^G & 0 & \Gamma_{23}^G \\ \Gamma_{31}^G & -\Gamma_{23}^G & 0 \end{pmatrix}, \tag{6}$$

with

$$\frac{\partial v_i^\alpha}{\partial r_i^\beta} = \sum_{j=1}^n \frac{m_j}{\rho_j}\left(v_j^\alpha - v_i^\alpha\right)\nabla_i^\beta W(\vec{\eta}_i)$$

$$\Gamma_{12}^G = \frac{\gamma_c\gamma_d - \gamma_b\gamma_e}{\gamma_a\gamma_c - \gamma_b^2}$$

$$\Gamma_{31}^G = \frac{\gamma_b\gamma_d - \gamma_a\gamma_e}{\gamma_a\gamma_c - \gamma_b^2}$$

$$\Gamma_{23}^G = \frac{G_{31}\Gamma_{12}^G + G_{21}\Gamma_{31}^G + G_{21}\sigma_{13}^G + G_{22}\sigma_{23}^G - G_{32}\left(\sigma_{22}^G - \sigma_{33}^G\right) - G_{31}\sigma_{12}^G - G_{33}\sigma_{32}^G}{G_{22} + G_{33}}$$

in which

$$\gamma_a = (G_{11} + G_{22})(G_{22} + G_{33}) - G_{31}^2$$

$$\gamma_b = (G_{22} + G_{33})G_{32} + G_{21}G_{31}$$

$$\gamma_c = (G_{11} + G_{33})(G_{22} + G_{33}) - G_{21}^2$$

$$\gamma_d = (G_{22} + G_{33})\left[G_{11}\sigma_{12}^G - G_{21}\left(\sigma_{11}^G - \sigma_{22}^G\right) + G_{31}\sigma_{32}^G - G_{22}\sigma_{21}^G - G_{32}\sigma_{31}^G\right]$$
$$+ G_{31}\left[G_{21}\sigma_{13}^G + G_{22}\sigma_{23}^G - G_{32}\left(\sigma_{22}^G - \sigma_{33}^G\right) - G_{31}\sigma_{12}^G - G_{33}\sigma_{32}^G\right]$$

$$\gamma_e = (G_{22} + G_{33})\left[G_{11}\sigma_{13}^G + G_{21}\sigma_{23}^G - G_{31}\left(\sigma_{11}^G - \sigma_{33}^G\right) - G_{32}\sigma_{21}^G - G_{33}\sigma_{31}^G\right]$$
$$- G_{21}\left[G_{21}\sigma_{13}^G + G_{22}\sigma_{23}^G - G_{32}\left(\sigma_{22}^G - \sigma_{33}^G\right) - G_{31}\sigma_{12}^G - G_{33}\sigma_{32}^G\right].$$

In the primary frame of the kernel, the $G$ tensor is read as

$$G^{(k)} = \begin{pmatrix} G_{11}^{(k)} & 0 & 0 \\ 0 & G_{22}^{(k)} & 0 \\ 0 & 0 & G_{33}^{(k)} \end{pmatrix}. \tag{7}$$

In general, $G_{11}^{(k)} \neq G_{22}^{(k)} \neq G_{33}^{(k)}$ in the ASPH. The full rotational transformation matrices $T_{\vec{r}}$ can be used to transform the $G^{(k)}$ tensor of the kernel primary frame to that of the real frame $G^{(r)}$. By choosing the roll angle $\phi$ about the x axis, the pitch angle $\psi$ about the intermediate y axis and the yaw angle $\theta$ about the z axis in the kernel frame, one can do this transformation from the following formula

$$G^{(r)} = \begin{pmatrix} G_{11} & G_{12} & G_{13} \\ G_{21} & G_{22} & G_{23} \\ G_{31} & G_{32} & G_{33} \end{pmatrix} = T_{\vec{r}}^T G^{(k)} T_{\vec{r}}, \tag{8}$$

where $T_{\vec{r}}^T$ is a transposed matrix of the $T_{\vec{r}}$ matrix and

$$T_{\vec{r}} = \begin{pmatrix} \cos\psi\cos\phi & \cos\psi\sin\phi & -\sin\psi \\ \sin\theta\sin\psi\cos\phi - \cos\theta\sin\phi & \sin\theta\sin\psi\sin\phi + \cos\theta\cos\phi & \cos\psi\sin\theta \\ \sin\theta\sin\psi\cos\phi + \sin\theta\sin\phi & \cos\theta\sin\psi\sin\phi - \sin\theta\cos\phi & \cos\psi\cos\theta \end{pmatrix}. \tag{9}$$

At here, one can set an initial one of the $G^{(r)}$ tensor by choosing $G_{11}^{(k)}$, $G_{22}^{(k)}$, $G_{33}^{(k)}$, $\phi$, $\psi$ and $\theta$, which present the shape of the particle ($G_{11}^{(k)}$, $G_{22}^{(k)}$ and $G_{33}^{(k)}$) and the orientation of the particle ($\phi$, $\psi$ and $\theta$).

In the present work, we use a Gauss kernel and its derivate as

$$W(\vec{\eta}_i) = \frac{\alpha_W^{3/4} |G_i|}{\pi \Gamma(3/4)} \exp(-\alpha_W \vec{\eta}_i^4), \tag{10}$$

$$\nabla W(\vec{\eta}_i) = -4\alpha_W W_i(\vec{\eta}_i) \vec{\eta}_i^2 G_i \vec{\eta}_i, \tag{11}$$

where $\vec{\eta}_i = G_i \vec{r}_{ij}$ and $\alpha_W$ is a parameter. The artificial viscosity function of the particle has the following form

$$\Pi_i = \begin{cases} \rho_i^{-1}(-\alpha_\Pi c_i \mu_i + \beta_\Pi \mu_i^2), & \vec{v}_{ij}.\vec{r}_{ij} < 0 \\ 0, & \vec{v}_{ij}.\vec{r}_{ij} \geq 0 \end{cases} \tag{12}$$

where $\vec{v}_{ij} = \vec{v}_i - \vec{v}_j$; $\alpha_\Pi$ and $\beta_\Pi$ are parameters; $c = \sqrt{\frac{h_{cr} p}{\rho}}$ is speed of sound of the particle, $p = (h_{cr} - 1)\rho u$ is pressure of the particle, $h_{cr}$ is a parameter; $\mu_i = \frac{\vec{v}_{ij}.\vec{\eta}_i}{\vec{\eta}_i.\vec{\eta}_i + \cdot^2}$, $\cdot$ is a very small parameter for avoiding numerical error. In all the calculations, $c_i$ and $\Pi_i$ are replaced by $c_{ij} = 0.5(c_i + c_j)$ and $\Pi_{ij} = 0.5(\Pi_i + \Pi_j)$, respectively. The stress tensor is

$$\sigma_i^{\alpha\beta} = -p_i \delta^{\alpha\beta} + S_i^{\alpha\beta}, \tag{13}$$

where $\delta^{\alpha\beta}$ is the Kronecker symbol and $S_i^{\alpha\beta}$ is the deviatoric stress calculated from the equation

$$\frac{dS_i^{\alpha\beta}}{dt} = 2\mu\left(\dot{\varepsilon}_i^{\alpha\beta} - \frac{1}{3}\delta^{\alpha\beta}\dot{\varepsilon}_i^{\gamma\gamma}\right) + S_i^{\alpha\gamma}R_i^{\beta\gamma} + S_i^{\gamma\beta}R_i^{\alpha\gamma}, \tag{14}$$

in which $\gamma \equiv x, y, z$, $\mu$ is the shear modulus of materials, the tensor of the rate of deformations $\dot{\varepsilon}_i^{\alpha\beta}$ and the tensor of stress rotation $R_i^{\alpha\beta}$ are read as

$$\dot{\varepsilon}_i^{\alpha\beta} = \frac{1}{2}\sum_{j=1}^{N}\frac{m_j}{\rho_j}\left[\left(v_j^\alpha - v_i^\alpha\right)\nabla_i^\beta W(\vec{\eta}_i) + \left(v_j^\beta - v_i^\beta\right)\nabla_i^\alpha W(\vec{\eta}_i)\right], \tag{15}$$

$$R_i^{\alpha\beta} = \frac{1}{2}\sum_{j=1}^{N}\frac{m_j}{\rho_j}\left[\left(v_j^\alpha - v_i^\alpha\right)\nabla_i^\beta W(\vec{\eta}_i) - \left(v_j^\beta - v_i^\beta\right)\nabla_i^\alpha W(\vec{\eta}_i)\right]. \tag{16}$$

In order to compensate energy dissipation caused by friction during the sliding, a dissipation force is afforced on each particle of the system as follows

$$F_{dis,i} = \begin{cases} -m_i\gamma_{dis}\left(v_i^x - V_{dis}\right) & \text{the x-direction} \\ -m_i\gamma_{dis}v_i^y & \text{the y-direction} \\ -m_i\gamma_{dis}v_i^z & \text{the z-direction,} \end{cases} \tag{17}$$

where $\gamma_{dis}$ is a parameter of the model and $V_{dis} = 0$ for particles of the substrate and $V_{dis} = V$, which is a constant sliding velocity of the slider, for particles of the slider. The Prandtl-Tomlinson model is utilized by adding a spring force on each particle of the slider as follows

$$F_{spr,i} = \begin{cases} K\left(x_{0,i} + Vt - x_i\right) & \text{the x-direction} \\ K\left(y_{0,i} - y_i\right) & \text{the y-direction} \\ K\left(z_{0,i} - z_i\right) & \text{the z-direction,} \end{cases} \tag{18}$$

where $K$ is a spring constant, $t$ is sliding time, $x_0$, $y_0$ and $z_0$ are the equilibrium/initial coordinates of the particle in the x-, y- and z-directions, respectively. Interaction between the slider and the substrate is presented by interaction between particles of the two layers, one of the slider and the other of the substrate, in the contact. Two particles, one of each layer, interact with each other by a spring force as follows

$$\vec{F}_{int,ij} = \begin{cases} -K_\alpha (r - r_{cut}) \dfrac{\vec{r}_{ij}}{r} & 0 < r \leq r_{cut} \\ 0 & r > r_{cut}, \end{cases} \quad (19)$$

where $K_\alpha$ is a spring constant and $r_{cut}$ is a cutoff of the force. The friction force $F_{fri}$, the normal force $F_{nor}$ and the friction coefficient $\mu_{cof}$ are defined as

$$F_{fri} = \sum_{i=1}^{N_f} \left( F_{spr,i}^x + F_{int,ij}^x \right), \quad (20)$$

$$F_{nor} = \sum_{i=1}^{N_f} \left( F_{spr,i}^z + F_{int,ij}^z \right), \quad (21)$$

$$\mu_{cof} = \frac{F_{fri}}{F_{nor}}, \quad (22)$$

where $N_f$ is the number of the particles (the friction particles) of the contact layer of the slider, $F^x$ and $F^z$ are the force components in the x- and z-directions, respectively.

## 2.2 Simulation system and parameters

A particle is created by lumping an atomic region of $n_x \times n_y \times n_z$ (number of the unit cells in the x-, y- and z-directions) oxide unit cells. This coarse-graining is similar to that done for $Cu_2O$ oxide [11]. Each CG particle is located at the center of mass of the corresponding

CG atomic region and has mass of $M_{CG} = n_x n_y n_z (3m_{Si} + 6m_O)$ with $m_{Si} = 28.085$ g/mol and $m_O = 15.999$ g/mol. This CG method converts the α-SiO$_2$ atomic system to a SiO$_2$ particle lattice system whose unit cell is characterized by three vectors $\vec{l}_a$, $\vec{l}_b$ and $\vec{l}_c$ with $\vec{l}_a = n_x a \vec{i}$, $\vec{l}_b = n_x b_1 \vec{i} + n_y b_2 \vec{j}$, $\vec{l}_c = n_z c \vec{k}$, $(\vec{l}_a, \vec{l}_b) = 120^0$, $(\vec{l}_a, \vec{l}_c) = 90^0$ and $(\vec{l}_b, \vec{l}_c) = 90^0$, where $\vec{i}$, $\vec{j}$ and $\vec{k}$ are the unit vectors of the three dimensional Cartesian-coordinate system in the x-, y- and z-directions, respectively; $a = 4.916$ Å, $b_1 = -2.458$ Å, $b_2 = 4.257381$ Å and $c = 5.4054$ Å [12]. A particle system of micronsize can be obtained from expansion of the particle unit cell along the three directions.

We consider $n_x = 1500$, $n_y = 1732$ and $n_z = 1364$ in this work, making the particle of the equal length in the directions $|\vec{l}_a| = |\vec{l}_b| = |\vec{l}_c| = 0.7374$ μm and its mass $M_{CG} = 1.061 \times 10^{-6}$ μg. The simulated particle system includes a 9000 particles slider of 33.7306 × 21.3839 × 6.6356 μm$^3$ and a 40000 particles substrate of 79.1151 × 36.1315 × 6.6356 μm$^3$ (Fig. 1). The initial distance between the slider and the substrate is set equally to $|\vec{l}_c|$. The lowest particle layer of the substrate is fixedly held during the simulations. The slider slides a constant velocity of 50 m/s in the x-direction. The parameters for α-SiO$_2$ are the density of 2.648 g/cm$^3$ and the shear modulus of 46.91 GPa [13]. The parameters for the ASPH are $\alpha_W = 0.1$, $h_{cr} = 1.4$, $\alpha_\Pi = 2$, $\beta_\Pi = 5$ and $\gamma_{dis} = 10^9$, $r_{cut} = \sqrt{2}|\vec{l}_a|$. The spring constant $K = 0.104$ N/m is taken from the spring constant of the α-SiO$_2$ oxide-functionalized cantilever [14]. The $K_\alpha$ spring constants are $K_x = K_y = K_z = 0.2K$, a ratio of 0.2 found to be well presenting spring sliding friction of CG microscale iron [15]. The ASPH simulation program is modified from the FDPS open code source developed by Particle Simulator Research Team (AICS, RIKEN, JAPAN) [16].

**3 Results and discussion**

Figure 2 shows the friction force and the normal force dependent on the sliding distance for the two initial shapes of the kernel or those of the particles $(G_{11}, G_{22}, G_{33}, \phi, \psi, \theta) = (0.2, 0.2, 0.1, 0, 0, 0)$ and $(0.2, 0.2, 0.05, 0, 0, 0)$ (in units of 1/µm and degree), $G_{ij} = 0$ with $i \neq j$; and the applied load of 10 µN. Each of these quantities are complete the same in the two behaviors. This indicates that spreading out the interaction region of the kernel or making a larger one of size of the particles in the normal direction does not result in the friction characteristics. Stick-slip friction regularly occurs during the sliding, with a longer one of the stick time than the slip time (the figure inserted in Fig. 2a). The ratio between the two types of the time is equal to 2.0. Most studies have reported a quickly slip duration, as a ratio of about 130 found in a study of the stick-slip phenomenon by scratching LiF single crystals with a diamond indenter [17]. However, the slip time is prolonged as temperature of the system increases, for example, ratios of about 9.0 and 2.3 at 25º C and 500º C, respectively, were experimentally observed in the stick-slip friction of CrVN; or 3.0 and 0.8 at 25º C and 500º C, respectively, for CrV(35%)N coatings [18]. The possible mechanisms for the prolonged slip duration were explained by forth and back jumps of the tips between two atomic positions before finally settling at the new position [19]. Our observations do not find dependence of this ratio on the particle shape. Due to interaction between the solid particles the exchange between the two states, the stick and the slip, does not sharply happen like that commonly found in friction of systems comprising interaction points. Interaction between the ellipsoidal particles can clearly be seen from the shape of the normal force curves (Fig. 2b).

The friction coefficient in the above behaviors oscillates in a steady interval from $\mu_{cof}^{\min} = 0.1228$ to $\mu_{cof}^{\max} = 0.1634$ leading to the average value of $\mu_{cof}^{ave} = 0.1376$ (Fig. 3a). These results are in a good agreement with friction coefficient of SiO$_2$ oxide mentioned previously experimental reports. The static friction coefficient and the dynamics friction coefficient of the clear quartz, the milky quartz and the rose quartz were found to be 0.11 and 0.10, 0.14 and 0.14, and 0.13 and 0.11 under the over-dried condition, respectively; or which are 0.11 and 0.10, 0.16 and 0.16, and 0.13 and 0.11 under the over-dried/air-equilibrated condition [20]. The silicon dioxide tip/silicon dioxide flat contact showed the static friction coefficients of $0.20 \pm 0.02$ and $0.15 \pm 0.02$ at the experimental conditions as the system in ultra-high vacuum ($\sim 5 \times 10^{-10}$ Torr) or Ar ($<10^{-6}$ Torr) and N$_2$ ($<10^{-6}$ Torr) [21]. As shown in Fig. 3b, the average friction coefficient ($\mu_{cof}^{ave}$) linearly decreases from 0.1379 to 0.1341 as the applied load increases in the range of 5 – 80 µN. Due to the very small reflection of the friction coefficient at the two ends of the applied load range, $\Delta\mu_{cof}^{ave} = 0.0038$, it can also be considered stably maintaining in this applied load range. The similar scenario for dependence of the friction coefficient on the applied load is seen in the study of Kumar et al. that showed that the dynamics friction coefficient of the fused silica sample derived with the 20 μm conical indenter is fixedly held at around 0.14 in the applied load range of 0 – 2000 μN [22]. It then linearly increases in the applied load range of about 2000 – 6000 μN [22]. It is worth noting that dependence of the friction coefficient on the applied load strongly varies with changing size of the indenter, a linearly rapid increase of the friction coefficient, from around 0.14 to 0.25, in the applied load range of about 0 – 750 μN with the 1 μm conical indenter [22]. The size of the slider in the present work is close to 20 μm in the x- or y-dircetion. Therefore, our result is in accordance with the experimental observation of Kumar et al.. The experimentally dynamic friction coefficient of the $SiO_2$/$SiO_2$ millimeter scale smooth contact was found to be decrease in the ultrahigh vacuum range of pressure of $10^{-2}$ – $10^{-7}$ Pa and increase in the ultrahigh vacuum pressure range of < $10^{-7}$ Pa [23]. However, the change of the friction coefficient was found to be small as seen whose values at some given pressures, the friction coefficients of 0.219, 0.176, 0.189, 0.198, 0.171, 0.157, 0.227 at the pressures of $10^5$, 10, 1, 0.1, $10^{-6}$, $5\times10^{-7}$, $3.6\times10^{-7}$ Pa, respectively [23]. As a summary, by using the ASPH simulations to monitor the sliding friction characteristics of the CG microscale $SiO_2$ oxide we find that the friction coefficient and the applied load – friction coefficient relationship show the similar values or the similar changes for the systems of size of micronmeter (our results), millimeter [21] and centimeter [20, 23]; especially, our simulation results are in an excellent agreement with the experimental observations [22] at the applied load of micron-Newton tens and the sliding object of microscale. The friction coefficient of $SiO_2$ oxide steady maintains with the applied low load, around micron-Newton tens.

Figure 4a shows dependence of the friction coefficient on the different orientations of the particles $(0,0,-30)$, $(0,0,30)$, $(0,0,-60)$, $(0,0,60)$, $(-45,-45,-45)$ and $(45,45,45)$ (degree) with the shape $(0.2,0.2,0.05)$ (1/μm) and the applied load of 10 μN. It is easy to see that the orientation of the particle almost does not result in the sliding time dependence of the coefficient due to coincidence of the curves. This indicates that both size and orientation of the kernel (or particle) in the ASPH do not influence on the characteristics of sliding friction of a stable system. Cause of this state can be explained by a little of varying of the kernel in a stable system, leading to the fact that interaction between particles is a little change during the sliding and the curves show a good periodicity. The ASPH have brought out the good results of the highly confused systems as the observations of change of the profile density [5] and the particle distribution [6]. Therefore, it can be nice to carry out calculations of tribological properties of a confused system including collision of rough contacts or asperities leading to form of debris and interaction between them. Such the problems will be aimed in our future works. It is also worth recalling that the ASPH still well presents sliding friction of a steady system, as proved in our present work. Figure 4b shows the change of the $G$ tensor via the sliding time for the two before-mentioned shapes $(G_{11}, G_{22}, G_{33}, \phi, \psi, \theta) = (0.2, 0.2, 0.1, 0, 0, 0)$ and $(0.2, 0.2, 0.05, 0, 0, 0)$ (1/μm, degree), $G_{ij} = 0$ with $i \neq j$; and the applied load of 10 μN. Value of each element at a given time is an average one as $1/N \sum_{I=1}^{N} G_{ij}^{I}$. The elements of the $G$ tensors vary very a little during the sliding. There is the same the scenario of the change of the $G$ tensor for the other behaviors (not shown). Because of the very small changes of the $G_{ij}$ elements, interaction between the particles is stably maintained during the sliding leading to the fact that the observed friction quantities are almost coincidence as shown the above.

Fig. 5 shows two of the simulated systems with the sinusoidal rough contact (Fig. 5a and 5b) and the simulation results (Fig. 5c and 5d) for the particle of $(0.2, 0.2, 0.05, 0, 0, 0)$ (1/µm, degree), $G_{ij} = 0$ with $i \neq j$; and the applied load of 10 µN. These systems are similar to sliding gouge models. It is clearly in Fig. 5c that at a given amplitude of the sinusoidal contact ($3|\vec{l}_c|$) the friction coefficient very slightly depends on the wavelength, taking a maximum value of 0.689 and an averaged value of 0.579 for all the cases. This result is in accordance with friction coefficient of quartz gouge. The steady state friction coefficient of quartz gouge has been founded to be 0.66 – 0.69 [24] and the order of 0.6 [25] in the experimental observations. In the experiments for montmorillonite – quartz simulated gouge, friction coefficient showed an increase as percentage of quartz increased and it reached a stable value of around 0.5 as quartz occupied from 85 – 100% [26]. Contrast to the wavelength dependence of the friction coefficient, the averaged friction coefficient shows a linear increase with the amplitude at a given wavelength, as seen in Fig. 5d. By extrapolating from the equation presenting the fitting line of the simulation data, the friction coefficient is founded to be 0.163 as the amplitude of the sinusoidal contact approaches to zero. This result is in a good agreement with the friction observations of the smooth contact discussed the above, where the averaged friction coefficient is 0.1376. Notice that the friction coefficient depends only the roughness of the contact, but not depend on changes of the particle shape and/or orientation because the $G_{ij}$ elements vary very a little during the simulations.

## 4 Conclusions

This work investigates dry sliding friction of the $SiO_2$ oxide by the coarse-grained particle model and the ASPH simulation. The simulation results in this work are in accordance with the experimental observations for the value of the friction coefficient (0.14) and the stability of the friction coefficient (around 0.14) in the applied low load range of micron-Newton tens. For a system fixedly held during the sliding, shape and orientation of the kernel or the particle almost do not result in the friction characteristics due to very a little varying of the kernel. For the sinusoidal rough contacts, at a given amplitude the friction coefficient almost does not depend on the wavelength while at a given wavelength it linearly increases with the amplitude.

**Figure captions**

**Fig. 1** The simulated particle system including the 9000 particles slider and the 40000

particles substrate.

**Fig. 2** The friction force (a) and the normal force (b) via the sliding distance with the two initial behaviors of the particle $(G_{11}, G_{22}, G_{33}, \phi, \psi, \theta) = (0.2, 0.2, 0.1, 0, 0, 0)$ and $(0.2, 0.2, 0.05, 0, 0, 0)$ (1/μm, degree).

**Fig. 3** The friction coefficient via the sliding distance (a) and the applied load (b) with the two initial behaviors of the particle $(G_{11}, G_{22}, G_{33}, \phi, \psi, \theta) = (0.2, 0.2, 0.1, 0, 0, 0)$ and $(0.2, 0.2, 0.05, 0, 0, 0)$ (1/μm, degree).

**Fig. 4** The friction coefficient via the sliding distance with the initial behaviors of the particle $(G_{11}, G_{22}, G_{33}) = (0.2, 0.2, 0.05)$ (1/μm) and $(\phi, \psi, \theta) = (0, 0, -30)$, $(0, 0, 30)$, $(0, 0, -60)$, $(0, 0, 60)$, $(-45, -45, -45)$ and $(45, 45, 45)$ (degree) (a) and the time evolution of the G tensor for the two initial behaviors of the particle $(G_{11}, G_{22}, G_{33}, \phi, \psi, \theta) = (0.2, 0.2, 0.1, 0, 0, 0)$ and $(0.2, 0.2, 0.05, 0, 0, 0)$ (1/μm, degree); $G_{ij}^t$ is a value at time $t$ (b).

**Fig. 5** Two of the simulated particle systems with a sinusoidal rough contact, $z = N_{am} |\vec{l}_c| \sin\left(\dfrac{2\pi y}{N_{wl} |\vec{l}_b|}\right)$, having the particle number of the slider, and of the substrate, $N_{am}$ and $N_{wl}$: 145440, 50040, 3 and 60 (a) and 146400, 48600, 9 and 40 (b). Dependence of the friction coefficient on the wavelength with $N_{am} = 3$ (c) and the amplitude with $N_{wl} = 40$ (d).

**Fig. 1**

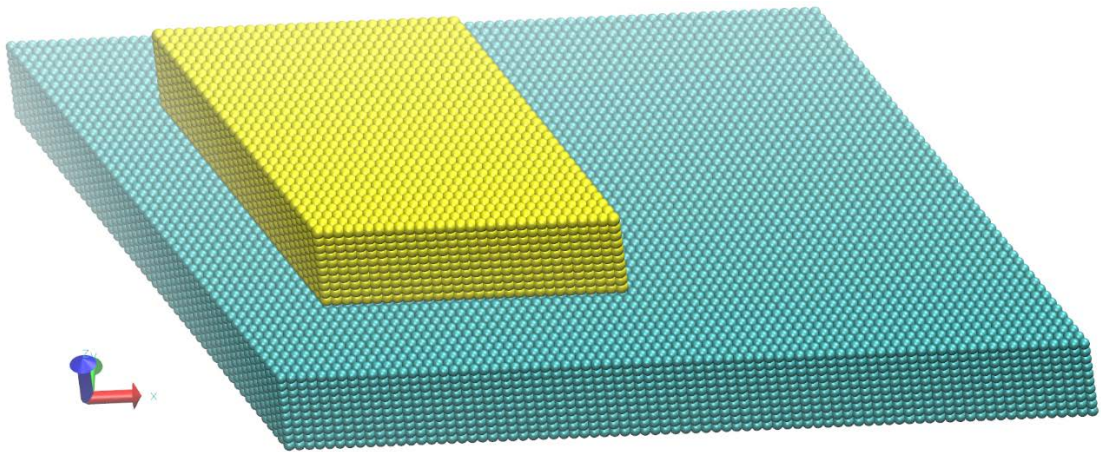

**Fig. 2**

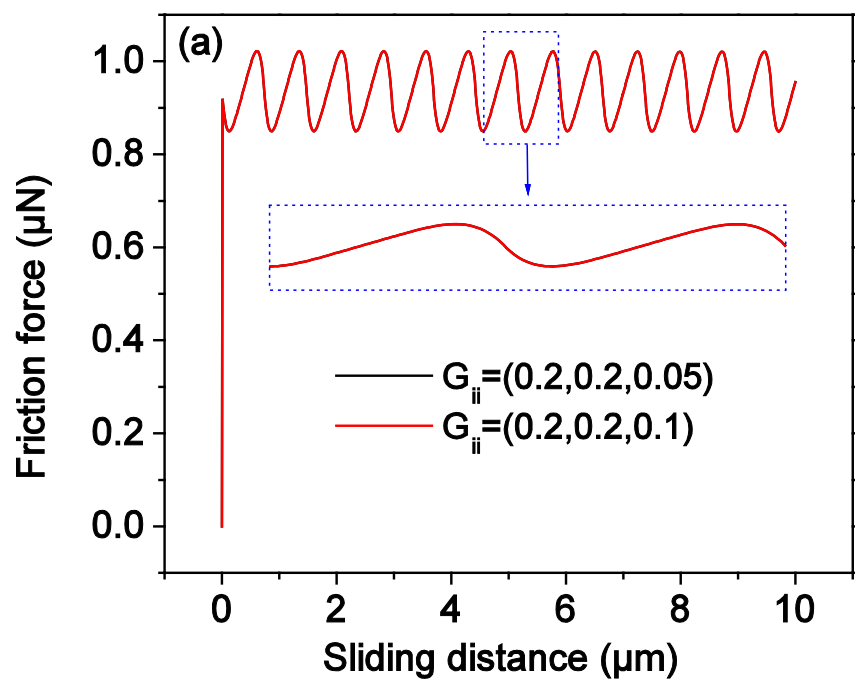

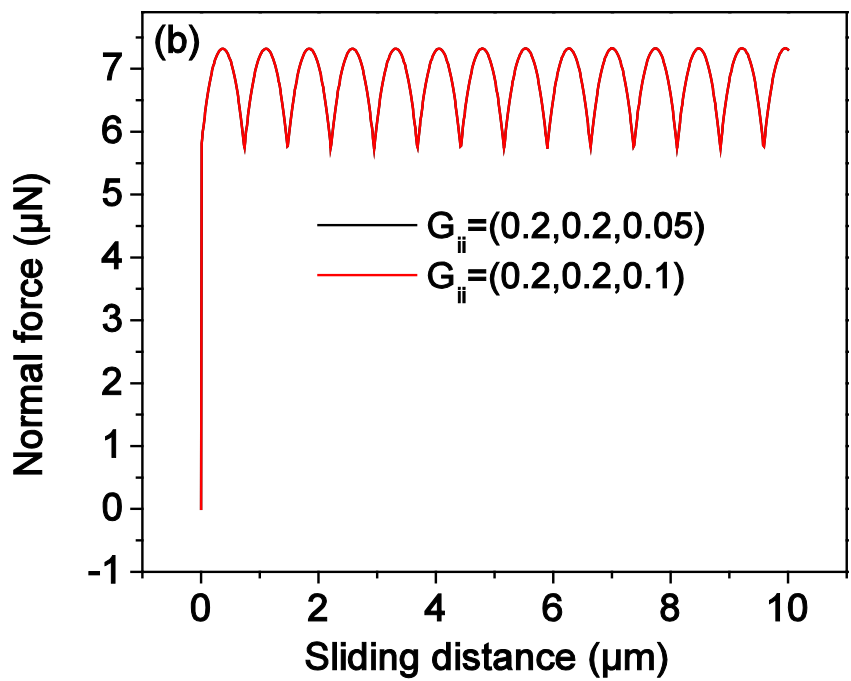

**Fig. 3**

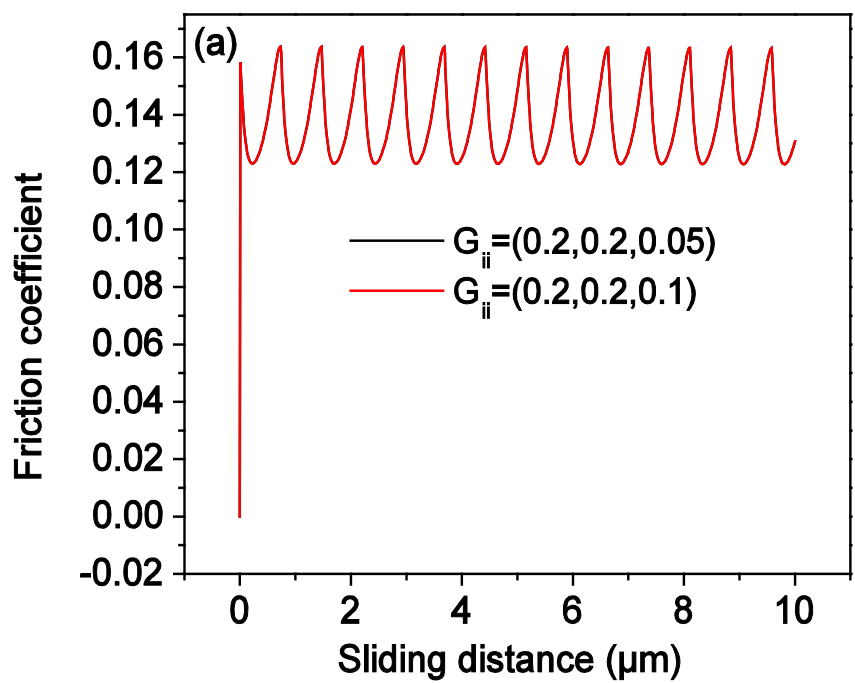

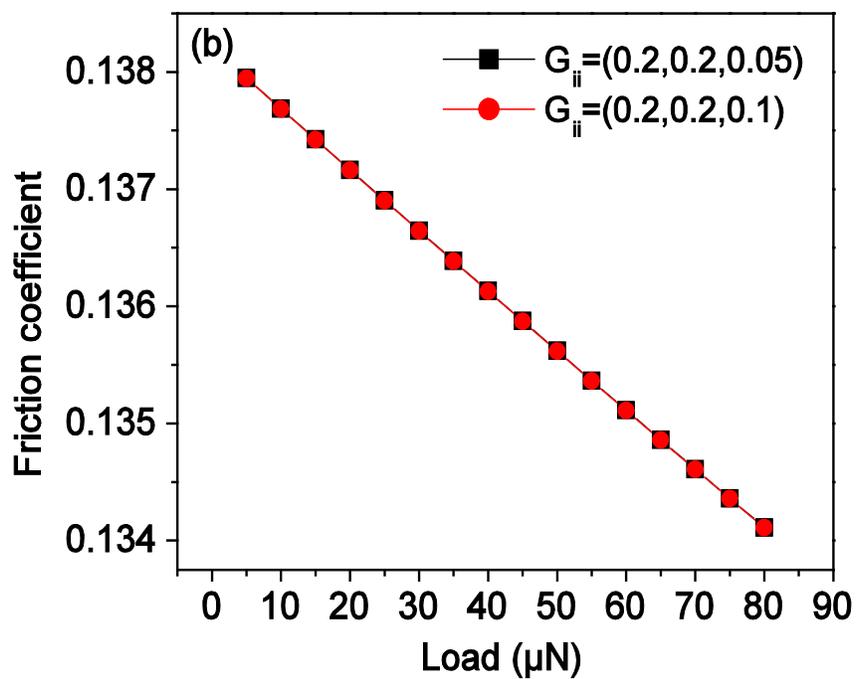

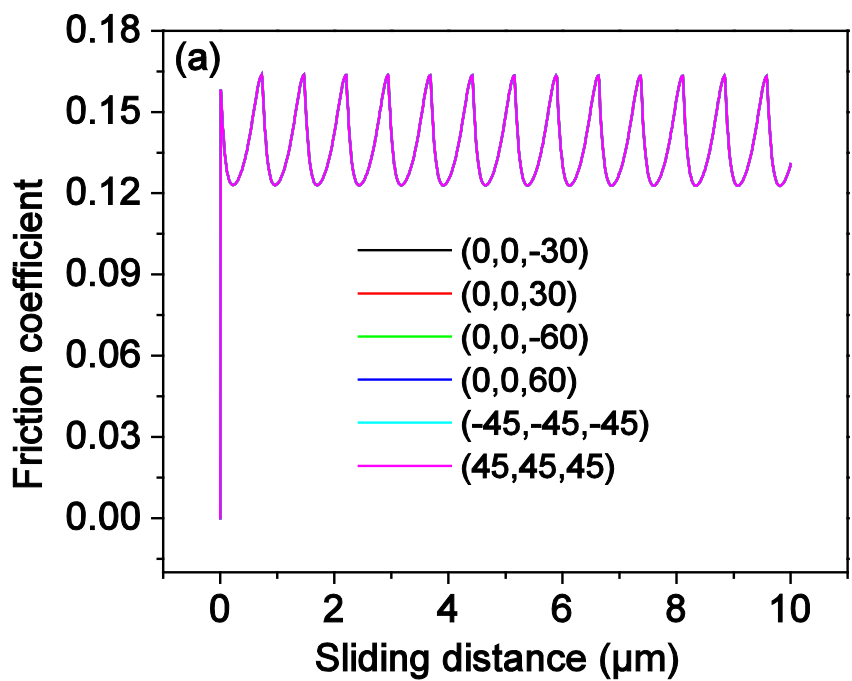

Fig. 4

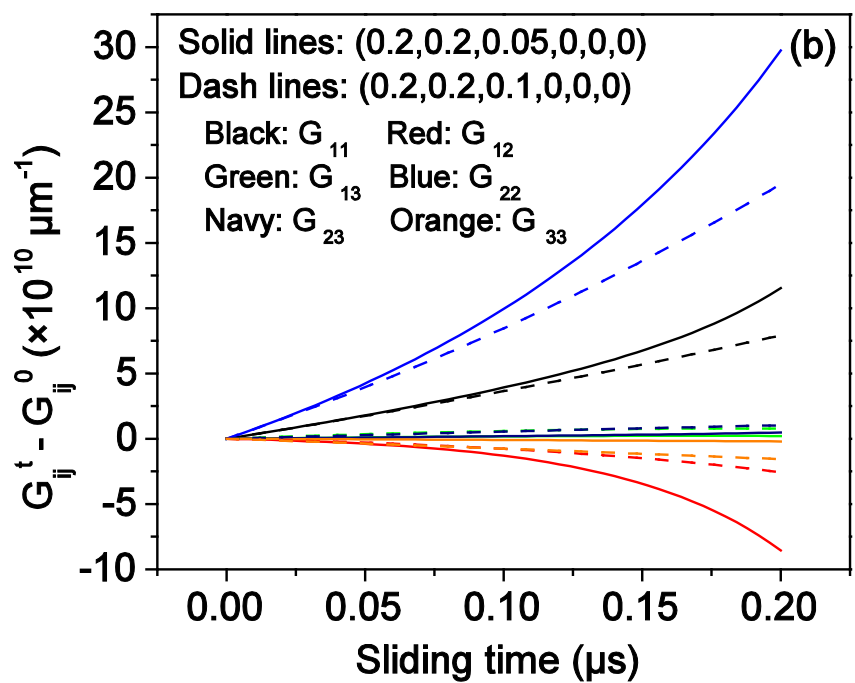

Fig. 5

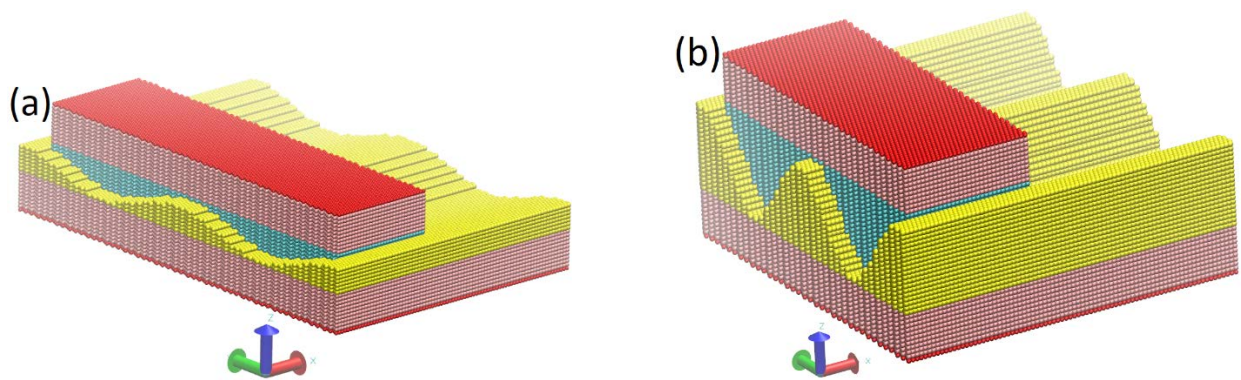

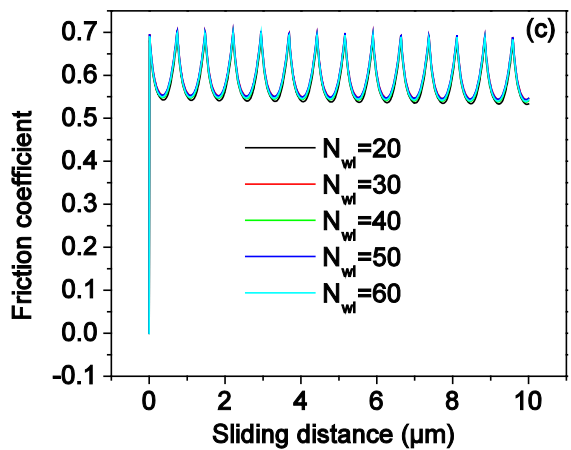 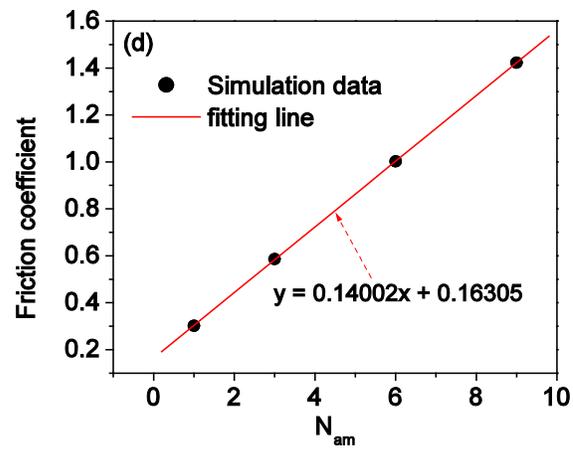

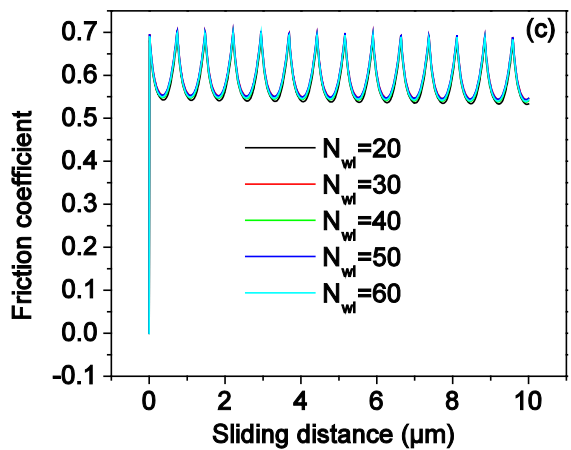 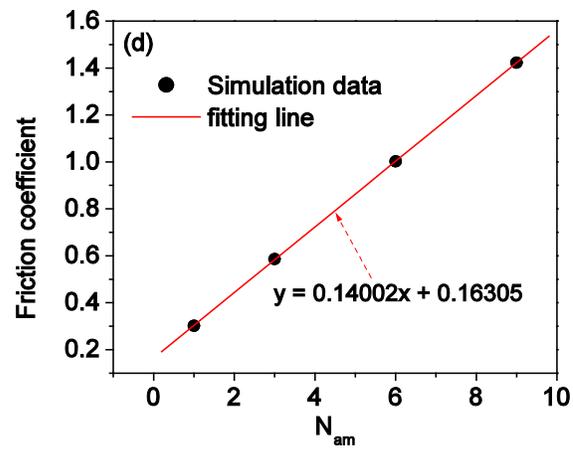